\pdfoutput=1
\documentclass[epjST]{svjour}
\usepackage{graphicx,url}
\usepackage[latin1]{inputenc}
\usepackage{listings}
\usepackage{graphicx}%
\usepackage{multirow}%
\usepackage{mathrsfs}%
\usepackage[title]{appendix}%
\usepackage{xcolor}%
\usepackage{textcomp}%
\usepackage{manyfoot}%
\usepackage{booktabs}%
\usepackage{algorithm}%
\usepackage{algorithmicx}%
\usepackage{algpseudocode}%
\usepackage{listings}%

\usepackage[font={small, sl}]{caption}

\usepackage{xspace}
\usepackage{color}
\usepackage{epsfig}     
\graphicspath{{.}{./figures/}}

\usepackage{keycommand} 
 \usepackage{hyperref}
  \usepackage{wrapfig}

\def\bR{\begin{color}{red}}  
\def\bB{\begin{color}{blue}}
\def\bM{\begin{color}{black}}  
\def\bC{\begin{color}{cyan}}
\def\bW{\begin{color}{white}}
\def\bBl{\begin{color}{black}}
\def\bG{\begin{color}{green}}
\def\bY{\begin{color}{yellow}}
\def\e{\end{color}\xspace}
\newcommand{\bit}{\begin{itemize}}
\newcommand{\eit}{\end{itemize}\par\noindent}
\newcommand{\ben}{\begin{enumerate}}
\newcommand{\een}{\end{enumerate}\par\noindent}
\newcommand{\beq}{\begin{equation}}
\newcommand{\eeq}{\end{equation}\par\noindent}
\newcommand{\beqa}{\begin{eqnarray*}}
\newcommand{\eeqa}{\end{eqnarray*}\par\noindent}
\newcommand{\beqn}{\begin{eqnarray}}
\newcommand{\eeqn}{\end{eqnarray}\par\noindent}

\begin{document}

\title{A Quantum Guitar}

\author{Bob Coecke\inst{1, 2, 3, 4, 5}\fnmsep\thanks{\email{bob.coecke@gmail.com}}}

\institute{
Quantum Brain Art Ltd, Oxford, UK \and 
Relational Intelligence Ltd, Oxford, UK \and 
Quantum Entertainment Berlin, Germany \and
Wolfson College, University of Oxford, UK \and 
Perimeter Institute, Waterloo, Ontario, Canada}
 
 \abstract{A guitar string represents a wave, and by associating a qubit to each of its playable states we get a quantum wave.  This is the principle behind our Quantum Guitar: `quantising all strings of a guitar'.  In order to achieve this quantisation, we couple a guitar with Moth's Actias quantum synth, and for qubit manipulations including measurements the musician uses foot controllers -- hence using all four limbs like a drummer. Our Quantum Guitar also allows for smooth continuous transition from `quantum' to `classical' sound, and vice versa.  We have used our Quantum Guitar in several live performances in a variety of venues, playing a number of different musical styles, hence demonstrating that it is a uniquely versatile and reliable  instrument.   For example, in some performances Quantum Guitar was used in industrial music, by the band Black Tish, who are currently recording an album with it. In other performances  Quantum Guitar represented a  qubit within a sonified Bell-pair under measurement, the other qubit being mentally realised with a Grand Piano. Classical-quantum and quantum-classical transitions prove particularly useful in musical performance. A link to a demo video of Quantum Guitar is provided.}
 
 
\maketitle

\section{Introduction} 

Some four years ago, for the first time, a piece of music was composed  with the use of a quantum computer, by a team under the pseudonym Ludovico Quanthoven \cite{quanthoven}, consisting of Miranda, Yeung, Pearson, Meichanetzidis, and Coecke. With the arrival of the field of quantum computer music more broadly \cite{miranda2022quantum}, there is a need for dedicated new `quantum musical instruments'.  These could be entirely new, or, could be the result of augmenting existing instruments.  The latter would enable one to quantum-enhance familiar modes of musical expression, that is, some familiarity of playing is retained. But, also,  radically new playing opportunities and entirely different sounding music should be aimed for.  The latter likely will require very different playing techniques, including new sophisticated coordinated uses of the musician's limbs. 

Different genres will require different instruments. By its very nature, one area of interest for electro-acoustic experimentation is Industrial Music \cite{reed2013assimilate,whittaker2022bodies}, a musical genre that has experimentation with sound  at its heart. One could say that Industrial Music is  the Musique Concr\`ete `of the people'.   Musique Concr\`ete, in the early 1940's, with composers like Schaeffer, Henry, Boulez, Stockhausen,  Var\`ese and Xenakis, made typically use of  pre-recorded unconventional sounds \cite{schaeffer2012recherche,schaeffer2017traite,kane2014sound}.  

Electrical guitars have been key to development of Industrial Music, with pioneering bands like Throbbing Gristle, Cabaret  Voltaire and Einst\"urzende Neubauten all employing guitar-alike instruments, as well as more experimental instruments, as in Musique Concr\`ete. 
\begin{figure}[h]
   \centerline{\includegraphics[width=170pt]{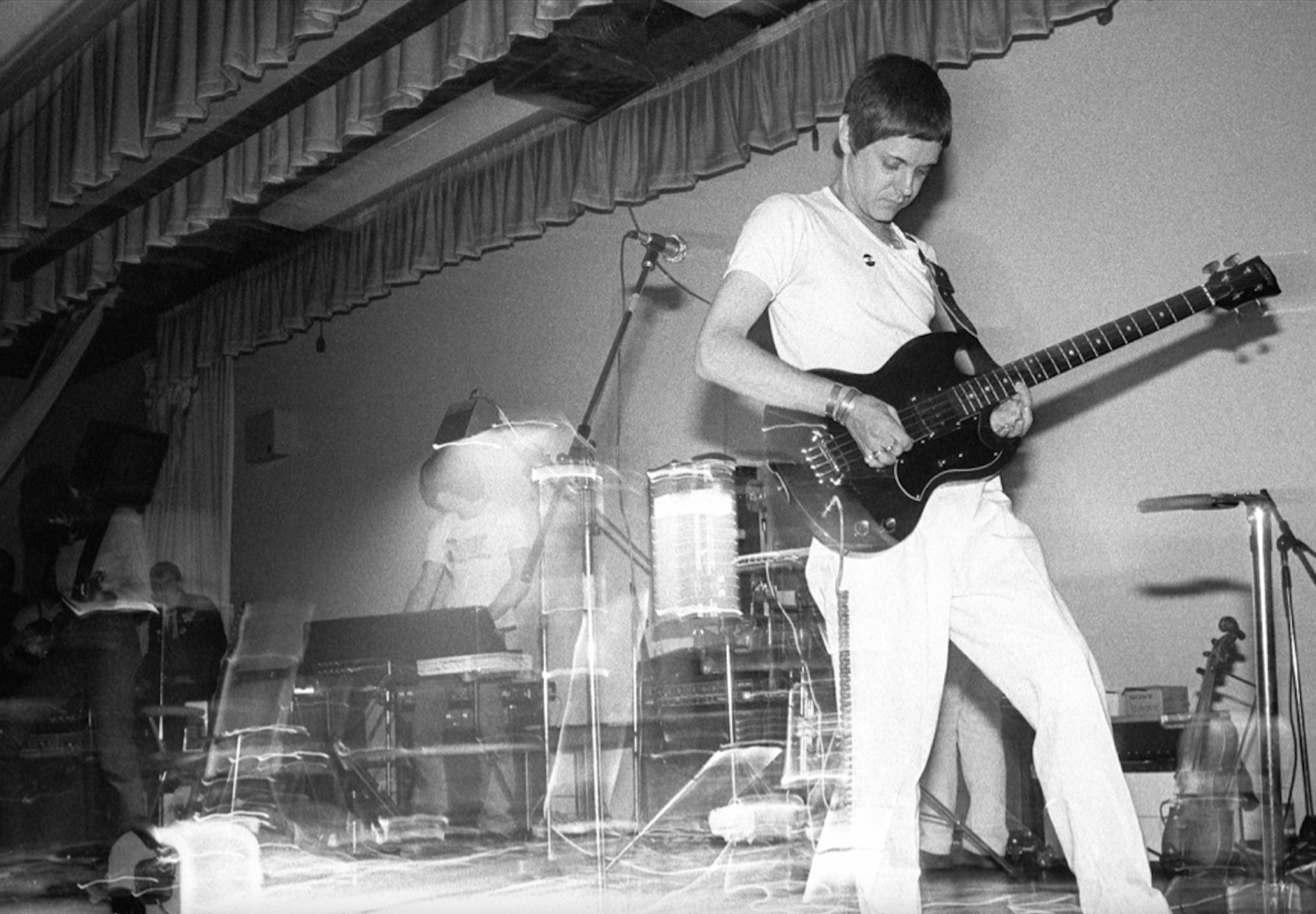}}
   \centerline{\includegraphics[width=170pt]{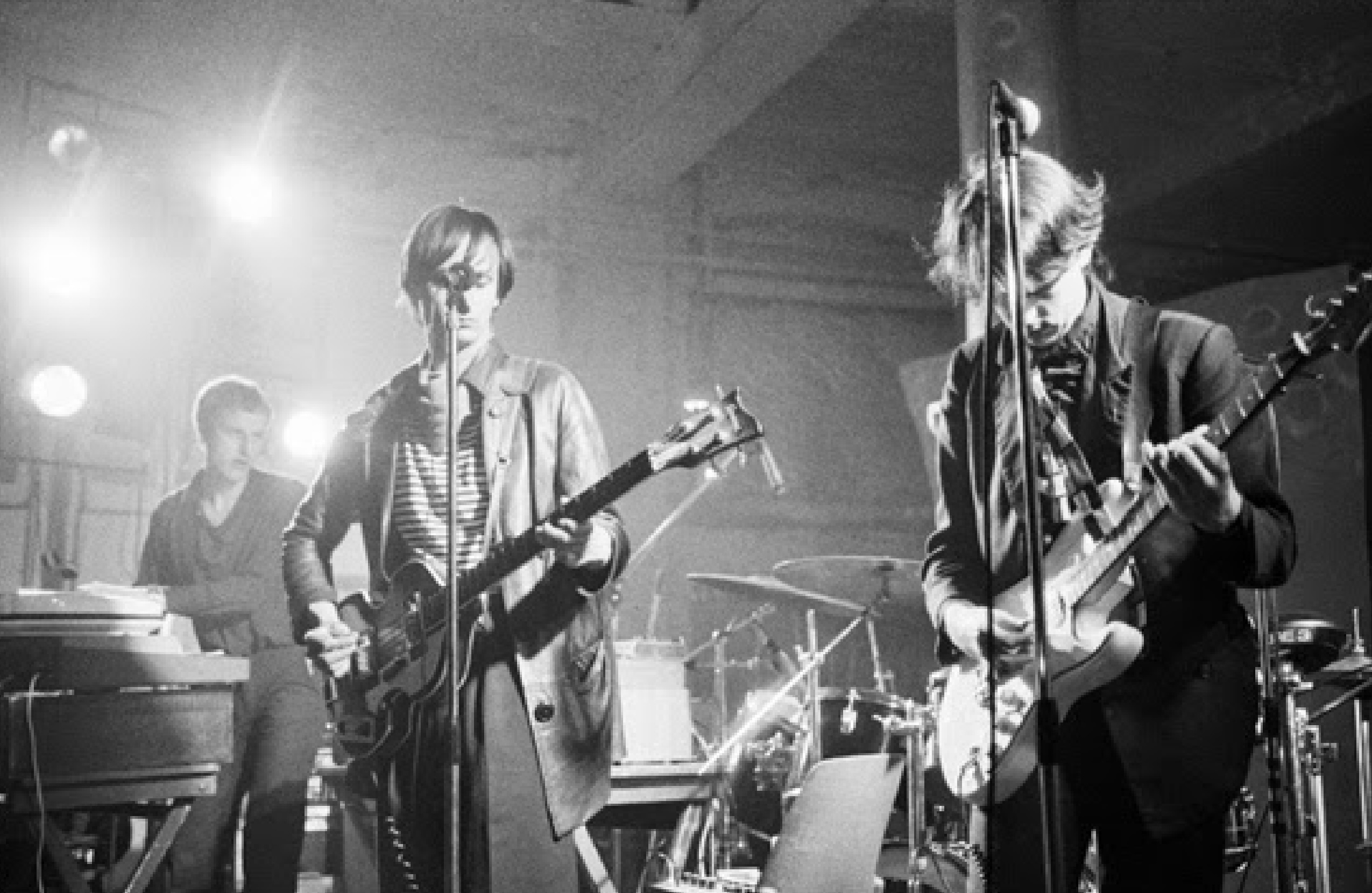}}
   \centerline{\includegraphics[width=170pt]{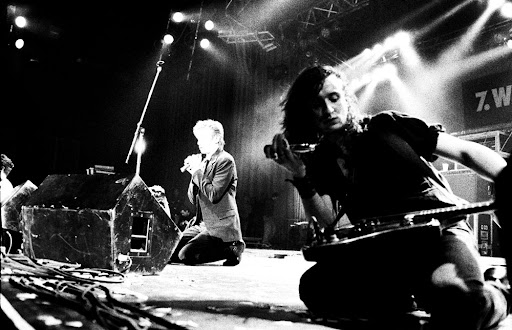}}
   \caption{Pioneers of Industrial like Music Throbbing Gristle, Cabaret  Voltaire and Einst\"urzende Neubauten all employed electrical (regular or bass) guitars.}
\end{figure} 

Currently Industrial Metal bands like NIN produce soundtracks for major Hollywood movies and have been introduced in the Rock'n Roll Hall of Fame.  Key ingredients of the genre have been integrated in virtually any musical style, such as metal, R\&B and rap, and all of electronic dance music is a direct descendent.  From  2024 onwards, the industrial band Black Tish started to incorporate quantum music, most notably in 2024 at Wacken Open Air, one of the (if not the) most prominent metal festivals in the world.  The band -- who retrospectively has been dubbed a pioneer of industrial music \cite{bigtakeover,youredm} --   is now recording an album entirely with Quantum Guitar.

What makes a guitar  unique among musical instruments is that the musician has the sound-generating substrate literally at their finger tips, which allows one to not only play with the traditional ingredients of music, but also exploit great scope for sound manipulation. For electrical guitar in particular, bends, overtones, damping, use of a tremolo bar, slide etc.~provide a wide spectrum for experimentation, especially when used with effect units that can emphasise and/or interact with these features, and therefore its broad use in Industrial Music.  

In this paper we describe a Quantum Guitar, that we have put together, and already used in several  live performances, \bM spanning heavy rock to contemporary classical music\e. These include academic events, major public science festivals, and most notably,  proper high-profile music venues, with the latter always posing an extra challenge.   The development went through a number of experimental phases, some which failed, and some which worked remarkably well.  We report here on the current architecture which, as just mentioned, already has proven itself as a battle-ready axe in a variety of circumstances.

A cornerstone for our Quantum Guitar is Moth's Actias, which is greatly inspired by the earlier Q1Synth \cite{miranda2023q1synth} due to Miranda, Thomas and Itabora{\'\i}.  However, previous uses never reached the full musical expression by a single musician that is now achievable by Quantum Guitar, in particular, how we manipulate the qubit.  

The first use of Quantum Guitar was at the Edinburgh Science Festival (Figure \ref{fig:Edinburgh}).  Other performances that have taken place paired  Quantum Guitar (played by ourselves) with Grand Piano (played by Rakhat-Bi Abdyssagin),  including performance of the  composition ``Bell" from \cite{BobRakhatI}, which uses a new augmented music notation based on ZX-calculus \cite{coecke2023basic}. 
\begin{figure}[h!]
   \centerline{\includegraphics[height=180pt]{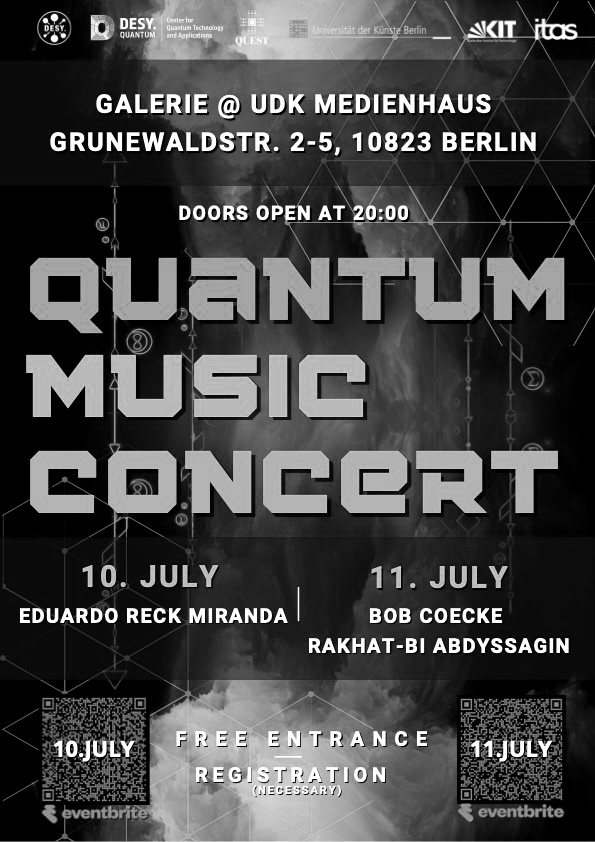}\includegraphics[height=180pt]{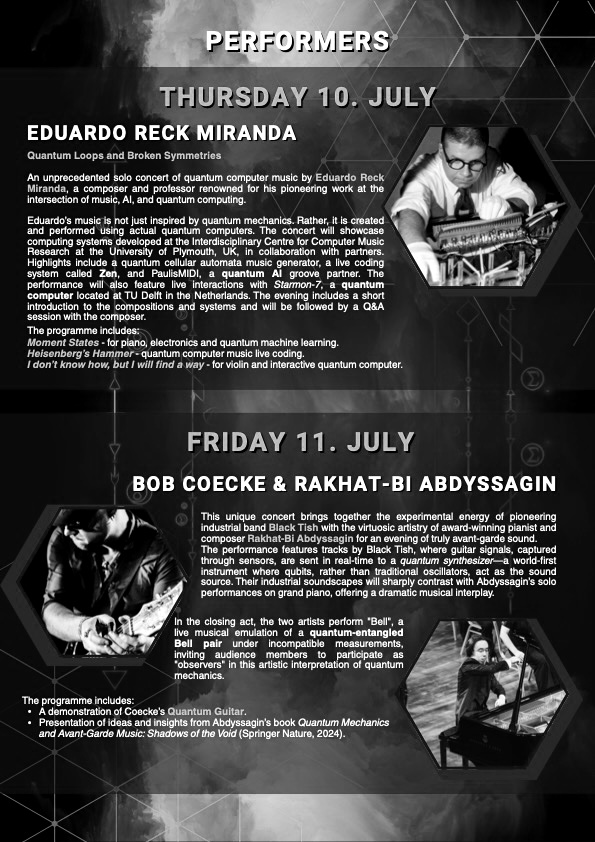}} 
   \caption{Poster from a performance in Berlin combining Quantum Guitar and Grand Piano.} 
    \label{fig:Berlin}
\end{figure} 
\bM The latest, and likely most high-profile performance, was the premiere of ``Quantum Universe", a symphony composed by Rakhat-Bi Abdyssagin for Cathedral Organ and Quantum Guitar, which is the first quantum symphony of its kind, performed in  Oxford's Merton College Chapel--see the video recording in Section \ref{sec:demo}.\e
Forthcoming performances include ``Quantum Universe" at Edinburgh's St Giles' Cathedral, and Quantum Guitar launch at Wacken Open Air metal festival, both August 2026. 
\begin{figure}[h!]
   \centerline{\includegraphics[height=250pt]{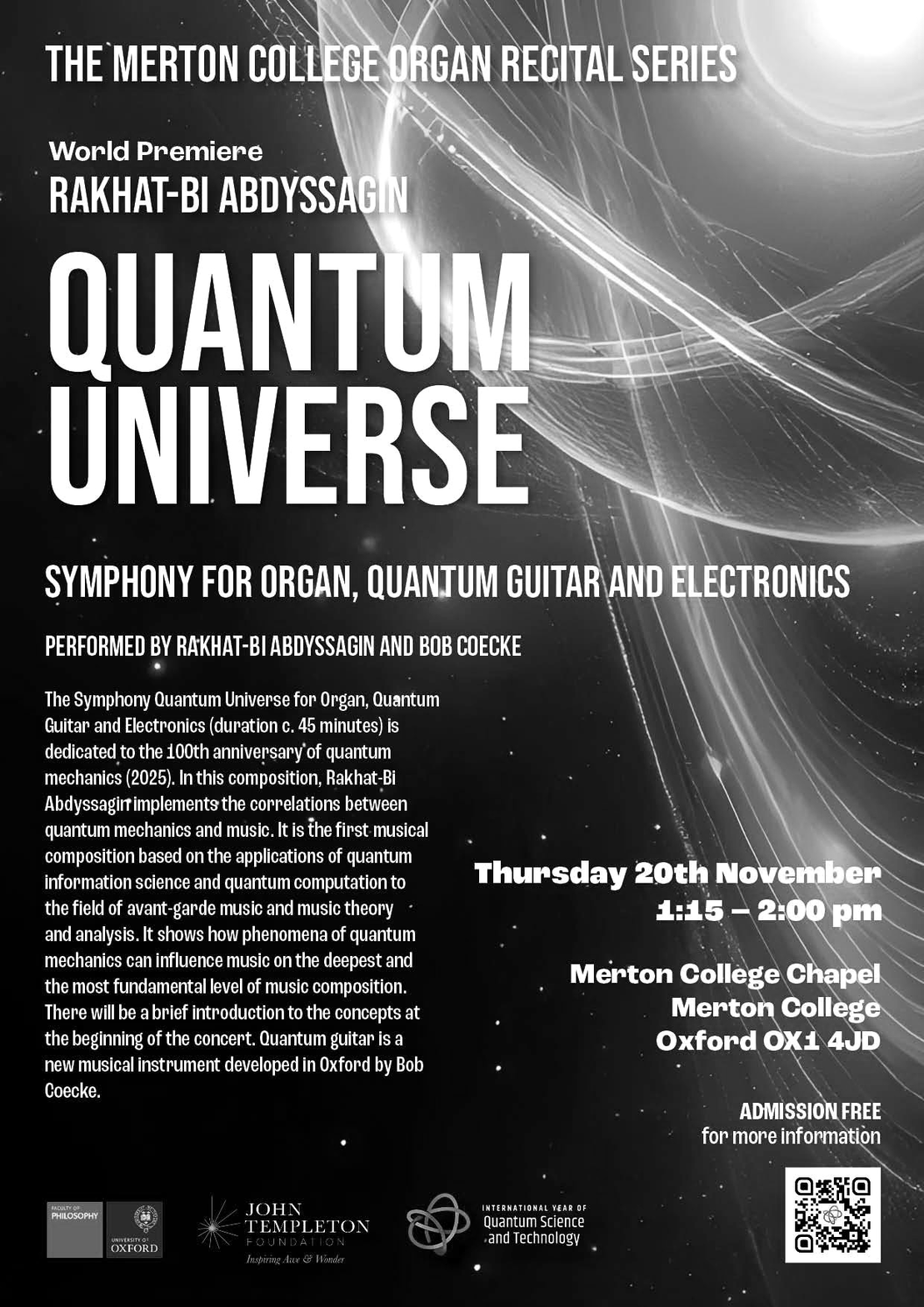}}
   \caption{\bM Poster of the premiere of the quantum symphony ``Quantum Universe".\e} 
    \label{fig:Merton}
\end{figure} 

 
\bM
\section{Physics semi-analogy}

Using a quantum physics analogy, and more specifically, quantum fields, the pitch constitutes the classical  spatial background, which quite literally, is the fret-board of the guitar. The idea of viewing a quantum field by associating a qubit to each spatial location is based on papers by Deutsch and others \cite{Deutsch2004,Pienaar2011,Singh2019}. Deutsch considers a field of identically-constituted quantum physical systems on spacetime, where each event is associated with one such system--and the simplest possible such system is a qubit. 

\begin{figure}[h!]
   \centerline{\includegraphics[height=120pt]{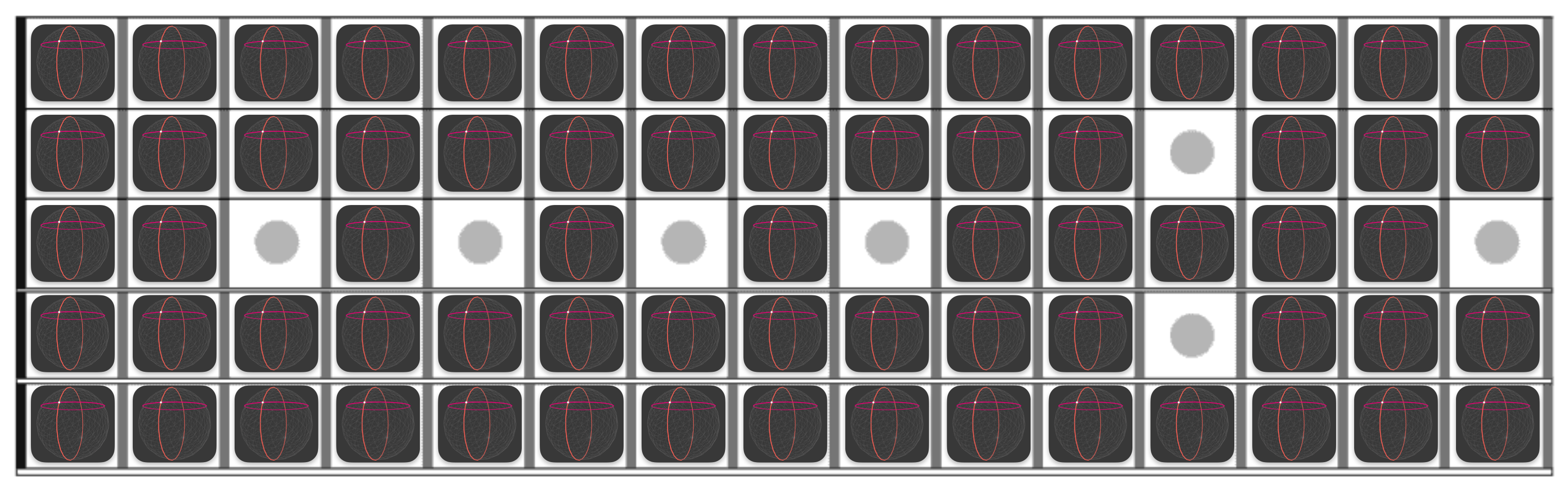}} 
   \caption{\bM Following Deutsch, a quantum field can be seen as identically-constituted quantum physical systems on spacetime, where each event is associated with one such system--and the simplest possible such system is a qubit.  Here space is the fretboard.\e} 
    \label{fig:Fretboard}
\end{figure}

The timbre is the results of the additional quantum degrees of freedom, that are provided by the associated qubit. So effectively, we quantise the fret-board of a guitar--see Fig.~\ref{fig:Fretboard}.  
The additional quantum variables provided by Quantum Guitar apply to all notes that are played, and we can change the associated qubit by changing those variables while changing pitch.  

Evidently, this is just the start of an implementation of such an analogy, and much more can be done.  As it stands, the fret-board is an analogy for physical space, while the qubit is an actual physical system, therefore,  `semi-analogy' in the title of this section.
At this point, of course, there is no direct analogy to particle-creation/destruction, but maybe that is a good thing.  

A formalisation of these ideas, and beyond,  is ongoing, in particular in the paper \cite{BobRakhatI} where categorical quantum mechanics \cite{AC1,CD2,ContPhys} seems to be the right place to do so, because of its trans-disciplinary nature \cite{DBLP:books/daglib/p/Coecke17}, with applications already existing in music \cite{quanthoven}, but also in natural language processing \cite{CSC,teleling}, in cognition \cite{ConcSpac},  in consciousness \cite{signorelli2020compositional}, and in several other domains \cite{CoeckeSpekkens2012,BaezLNP}, in particular all of those that can be cast as resource theories \cite{CFS}.
\e

\section{Demo and concert videos}\label{sec:demo}

A demo of Quantum Guitar recorded at my home studio (using Fender EOB with sustainer pick-up) is available at:
\beq\label{eq:blackpub}
 \href{https://www.youtube.com/watch?v=Pr4Wr8fdsL0}{https://www.youtube.com/watch?v=Pr4Wr8fdsL0} 
\eeq
\beq
 \href{https://zenodo.org/records/22101117}
 {https://zenodo.org/records/22101117} 
\eeq
\bM A concert video involving quantum guitar recorded at Merton College Chapel (using Ibanez KRYS10 with Mercury X reverb) is available at:
\beq\label{eq:merton}
 \href{https://youtu.be/1Dl8BGqTuE0}{https://youtu.be/1Dl8BGqTuE0} 
\eeq
We discuss the use of these particular electric guitars below.\e

\section{High-level design}

The basic idea of Quantum Guitar is as follows:
\ben
\item[(1)] A guitar string represents a wave, and by associating a qubit to each of its playable states we obtain a quantum wave, i.e.~quantisation of a guitar by means of quantising its strings.
\item[(2)]  As all hands/fingers are needed for ordinary guitar playing,  in order to control the extra degrees of freedom that come with this quantisation the musician uses their feet, just like a drummer uses all four limbs.  
\een
Moreover, one important aspect of this quantum instrument that has been somewhat overlooked in the past for other quantum instruments is:
\ben
\item[(3)]  One can  transition continuously  between quantum and classical sound, the latter being ordinary electrical guitar.
\een
Just as it is the case for a drummer, Quantum Guitar requires sophisticated coordination of four limbs, and hence is a new instrument that requires substantial playing practice for adequate artistic exploration of the augmented sound space.  The musician evidently needs to be seated for the performance in order to use both feet at the same time when manipulating the qubit, or for performing classical-quantum transition. 

Here are the key high-level ingredients of our Quantum Guitar:
\bit 
\item a MIDI controlled quantum synth -- sonify qubit
\item a MIDI-pickup --  generate MIDI data 
\item MIDI foot controllers -- perform qubit rotations
\item MIDI foot switches -- perform measurements
\item two volume control pedals -- classical vs.~quantum
\eit
In addition, in order to overcome the limitation of a guitar's relatively short-lived notes:
\bit
\item a sustainer pickup for generating continuous quantum sound.
\eit
Alternatively:  
\bit
\item a good quality delay pedal and/or a high-end reverb pedal could achieve this as well, and adds an extra creative sonic element.
\eit

\section{Concrete implementation} \label{sec:concrete}

\begin{figure}[h!]
   \centerline{\includegraphics[height=220pt]{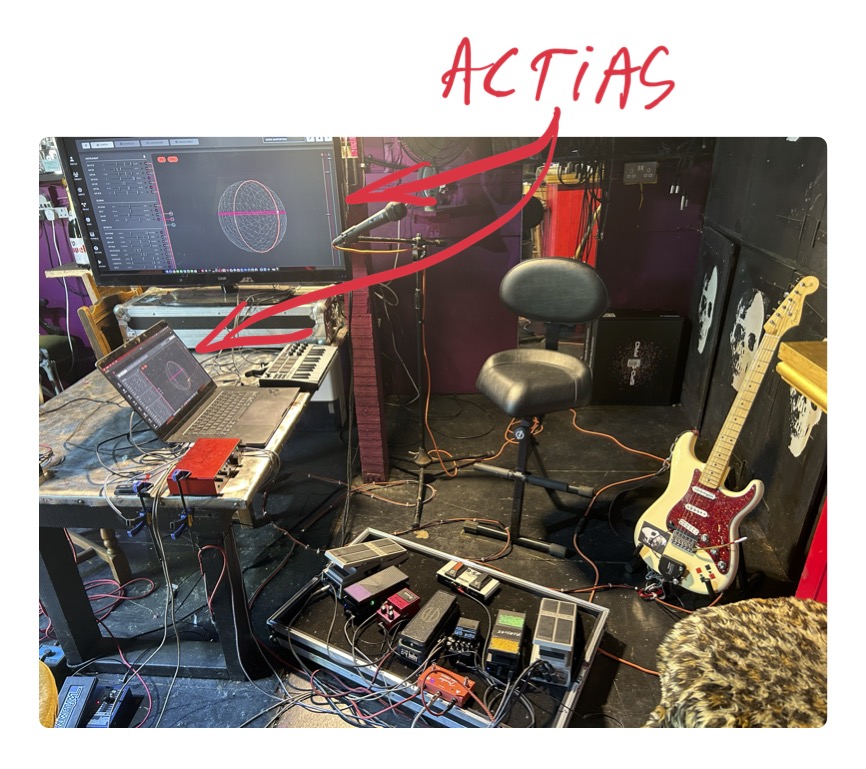}}
   \caption{View of our Quantum Guitar setup, using Actias for sonification.}
    \label{fig:Actias}
\end{figure} 
\begin{figure}[h]
   \centerline{\includegraphics[height=110pt]{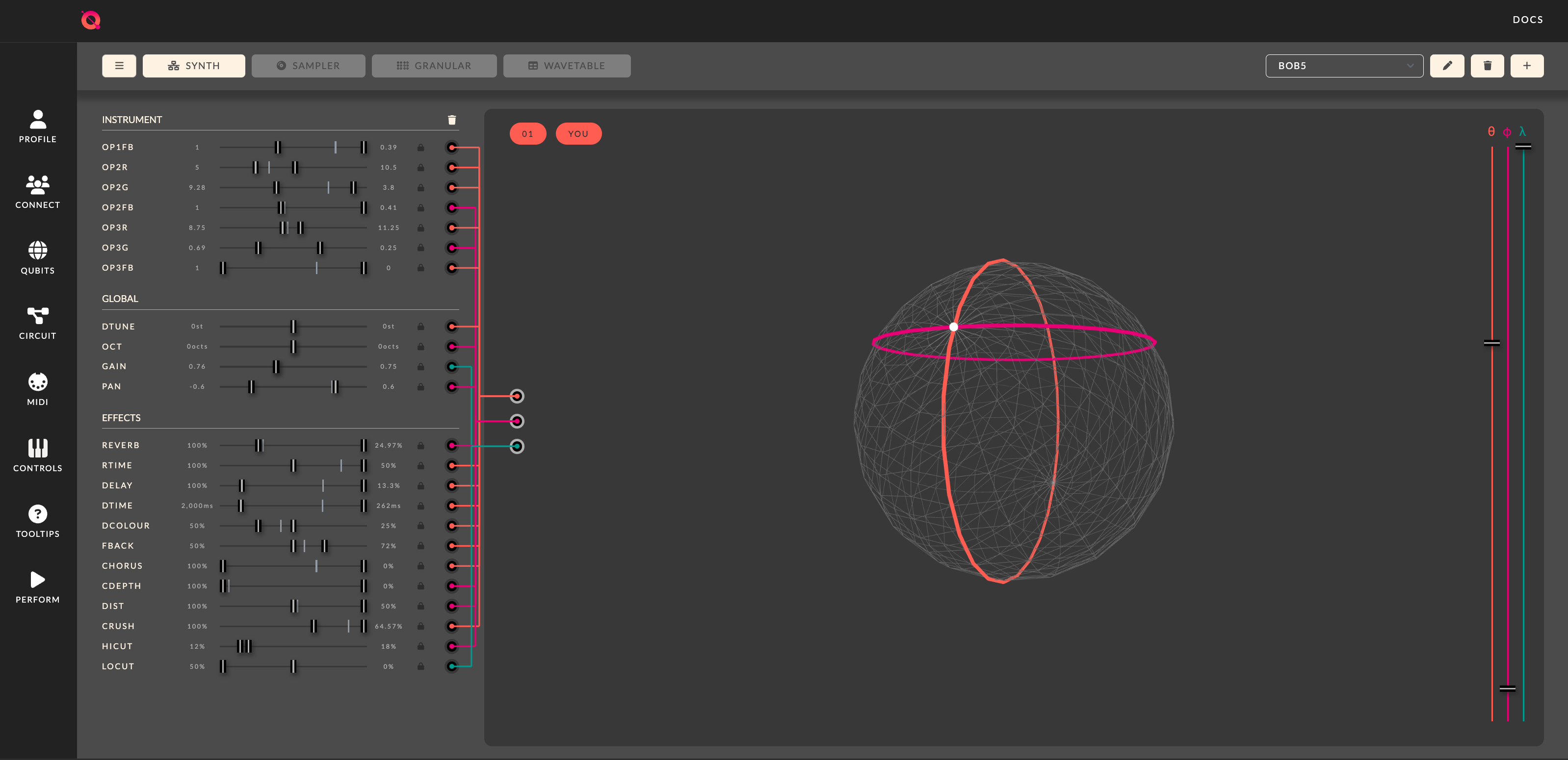}}
   \caption{The qubit depicted in Actias.}  
    \label{ActiasQubit}
\end{figure} 
\begin{figure}[h!] 
   \centerline{\includegraphics[height=260pt]{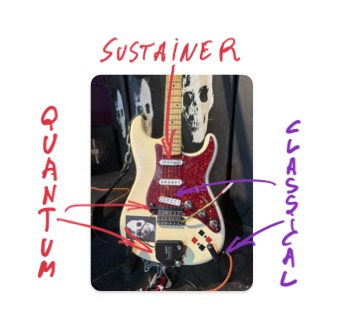} }
   \caption{Detailed Quantum Guitar setup part I.  The guitar is equipped with a high-quality reliable MIDI-pick-up (Fishman) and a sustainer pick-up that generates long maintainable notes (Fernandes). The guitar both generates `classical' and `quantum' sound at the same time.}
   \label{fig:guitar}
\end{figure} 
\begin{figure}[h!]
   \centerline{ \includegraphics[height=220pt]{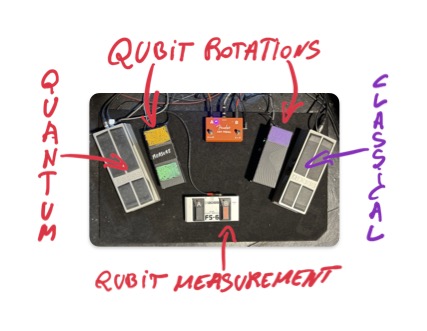}}
   \caption{Detailed Quantum Guitar setup part II.  Reliable MIDI foot controllers (Boss) and a foot switch are used in order to rotate and measure the qubit.  Volume pedals are used to independently control the `classical' and the `quantum' sound. One can recognise the colours corresponding to the Actias display on the two MIDI controllers, which control the corresponding rotations.  In this setup, ``green" and ``orange" cannot be used independently.}
\label{fig:pedals} 
\end{figure} 
The devices we use are (see Figures \ref{fig:Actias}, \ref{ActiasQubit}, \ref{fig:guitar} and \ref{fig:pedals}):
\bit
\item We employ Moth's Actias synth for sonification of the qubit. This allows rotation of the qubit in any possible manner, as well as one measurement of the qubit.  Obviously Moth's Actias software is experimental, and that comes at a price, as we discuss below.
\item We use a Fishman MIDI pickup system that has proven to be very good quality, very reliable and works well enough with Actias. Given the continuous spectrum of frequencies and other features of guitar playing, one needs a very  good quality MIDI pickup that doesn't miss out on `note off's', and translates playing with bends and tremolo bar appropriately using pitch bends. 
\item We use Boss EV-1-WL MIDI expression pedals to manipulate the qubit.  They also prove to be very reliable and work well with Actias, especially when used with a USB cable.  While we only use two pedals, we can also cover the third qubit rotation needed for all qubit manipulations, due to and internal switch feature of the Boss EV-1-WL pedal.  The use of only two pedals  is justified by the fact that we are working with a single qubit, so ``green" (see Figure \ref{fig:pedals}) is not supposed to change the state.  For demonstration purposes it is still good to have ``green" at hand.
\item For measurement we  use a Boss foot-switch like FS-6 in `momentary mode' -- `latch mode' won't work.  
\item We use Boss  FV500L/H volume pedals since their increased size and weight allows for subtle precision.
\item For obtaining a continuous sound, we use a Fernandes sustainer pickup, as shown in Picture \ref{fig:guitar}.  Alternatives that we have explored and also work are long delays and high-end reverbs with `infinite' capability, \bM applied to the audio output of the quantum synth.  One such setup that we  regularly use involves the (amazing) Mercury X Reverb\e.
\eit

\begin{figure}[h!]
   \centerline{\includegraphics[height=480pt]{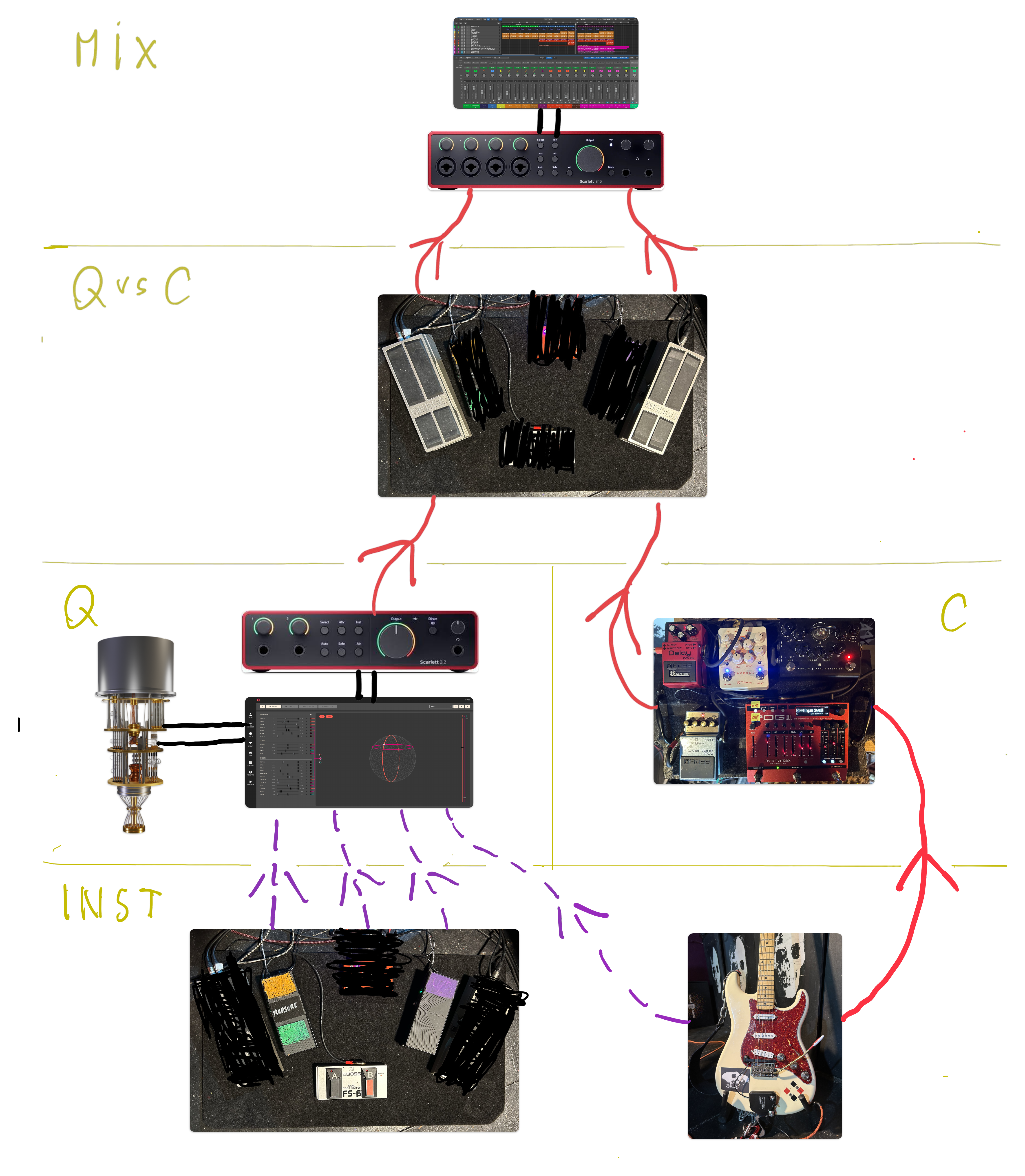}}
   \caption{\bM Pipeline for Quantum Guitar ideal for live performance.  An alternative setup  
that may be more useful for recording and composing, would consist of sending the MIDI data first to Logic Pro, and from there to Actias, as it allows for the MIDI to be directly edited before Actias converts it into audio.    This also requires   MIDI-interfaces for each of the laptops, so they can be connected via MIDI.\e} 
   \label{fig:pipeline}
\end{figure}

\bM Figure \ref{fig:pipeline} gives the full pipeline of the setup.  We can see five parts of the pipeline, from bottom to top:
\bit
\item {\bf INST:} The `instruments' layer consisting of the (1, 2) {\bf MIDI pedals} and (3) {\bf switch} on the left, and the (4) {\bf electrical guitar} with mounted (5) {\bf MIDI-pickup}. 
\item  {\bf Q:} The `quantum processing' layer consisting of a (6) {\bf laptop with Actias} that connects to an external (7a) {\bf quantum computer} or an internal (7b) {\bf quantum simulator}, and via USB connects to an (8) {\bf audio interface}.
\item  {\bf C:} The `classical processing' layer consisting of a (9) {\bf guitar pre-amp} and desired additional guitar pedals. 
\item {\bf Q vs C:} The `quantum versus classical balance' layer consisting of  (10, 11) {\bf two audio volume pedals}, one for the quantum audio signal and one for the classical audio signal.  Evidently, this also allows for controlling the overall volume. 
\item {\bf Mix:} The `mixing desk' layer consisting of (12) {\bf laptop with Logic Pro} that via USB connects to a second (13) {\bf audio interface}.
\eit
The wirings should be clear from Figure \ref{fig:pipeline}. The full lines represent audio, the dashed lines represent MIDI, and the double lines represent computer data connections, e.g.~USB, and are bi-directional. 
Personally I use a wireless connection between the MIDI-pickup and the laptop which has been proved to be very stable for my choice of MIDI-pickup.

An alternative setup 
that may be more useful for recording and composing, would consist of sending the MIDI data first to Logic Pro, and from there to Actias, as it allows for the MIDI to be directly edited before Actias converts it into audio.    Besides the audio interfaces:  
\bit
\item then also {\bf two MIDI-interfaces} are needed, one for each of the laptops, so they can be connected via MIDI. 
\eit
Ideally these are part of the audio interface in order to reduce cables and avoiding extra troubleshooting.

\begin{figure}[h!]
   \centerline{\includegraphics[height=120pt]{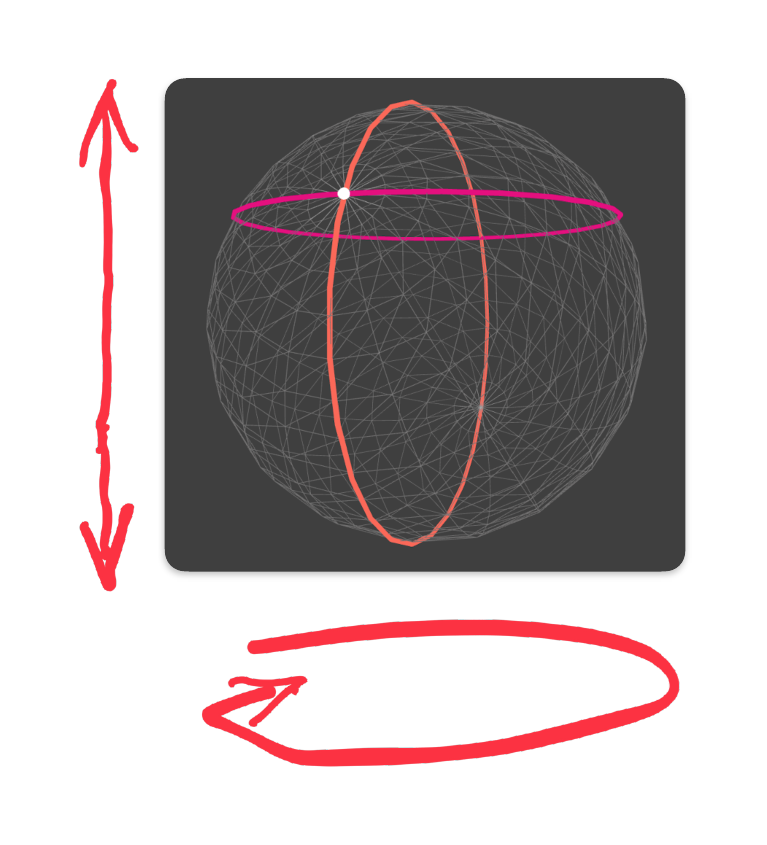}}
   \caption{\bM Parametrisation of a qubit state space in Actias: rotation around the Z-axis and translation along the Z-axis.\e} 
   \label{fig:para}
\end{figure}

\begin{figure}[h]
   \centerline{\includegraphics[height=160pt]{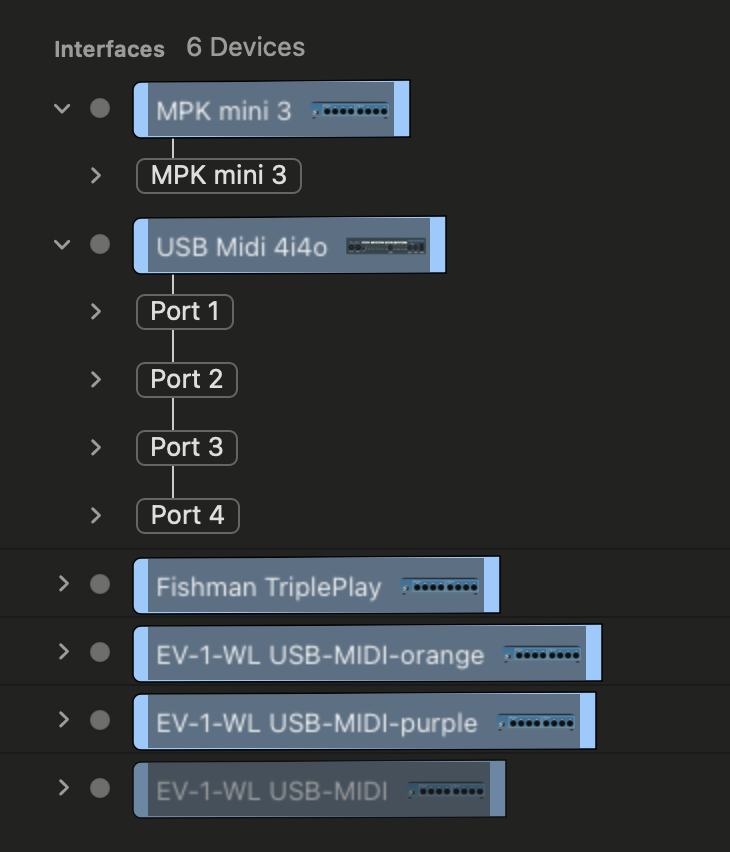}\ \ \ \ \includegraphics[height=160pt]{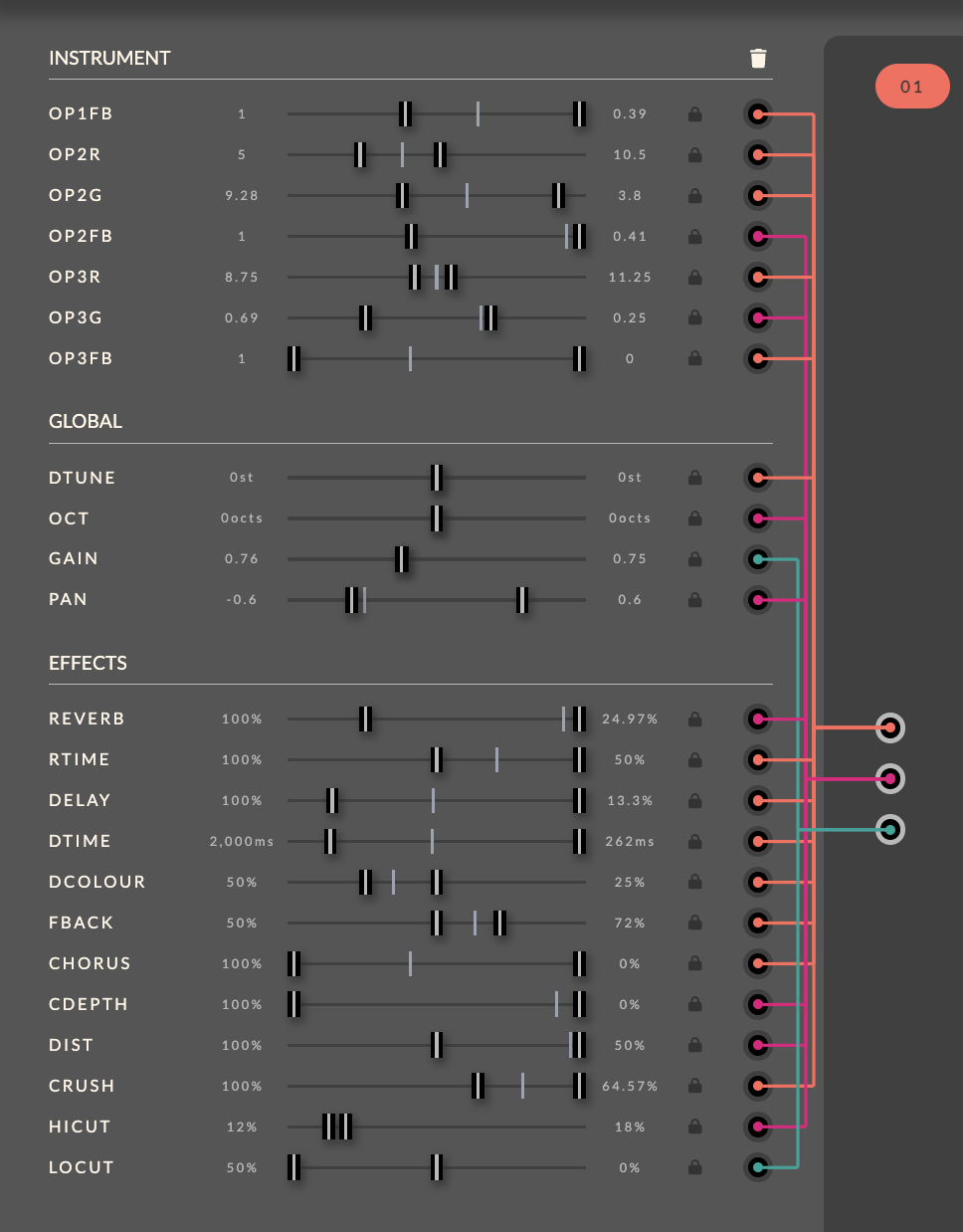}}
   \caption{\bM Our Mac MIDI setup and parameter settings for Actias.\e} 
   \label{fig:ActiasMIDI}
\end{figure}

Zooming in specifically on the use of Actias, technically, it operates through Mac's built-in MIDI setup--see Figure \ref{fig:ActiasMIDI} for my setup.  The code as well as installation instructions is available here:
\begin{center}
 \href{https://github.com/moth-quantum/actias-electron}{https://github.com/moth-quantum/actias-electron} 
\end{center}
\begin{center}
 \href{https://github.com/moth-quantum/actias-backend}{https://github.com/moth-quantum/actias-backend} 
\end{center}
For easy use, one can use Moth's internal quantum simulator, which, as far as one qubit is concerned, is perfectly adequate.  The sonification process then involves setting a number of variables in the control panel--see Figure \ref{fig:ActiasMIDI} for my settings.  In the latter, one can see the three dots on the right, which correspond to the parametrization of the Bloch sphere, and for each parameter one can assign settings for the quantum synth.  In addition, they can control also some global sound settings, as well as some additional effect settings.  The exact nature of the variables can be best seen in the demo of Section \ref{sec:demo}.  One of these parameters is rotation around the Z-axis,  the second one is translation along the Z-axis, and the third one is rotation along the axis of the qubit state--Fig.~\ref{fig:para}.  Evidently, the latter doesn't change the state, except for when the qubit would be entangled.  Therefore we typically don't use it in performances, in order to be loyal to the physics of a single qubit.
\e

\begin{figure}[h]
   \centerline{\includegraphics[height=240pt]{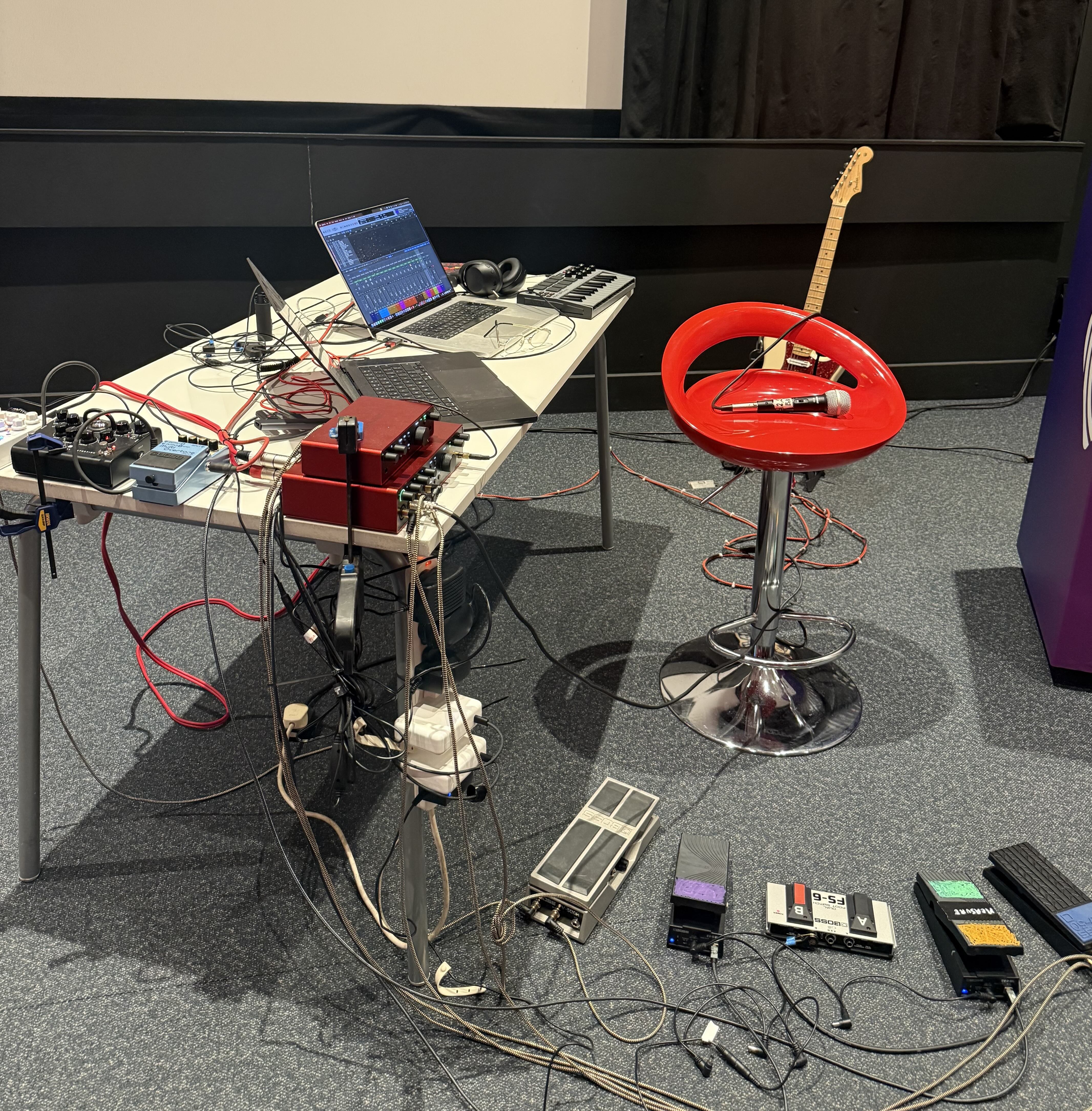}}
   \caption{Quantum Guitar setup for a live performance 
   at Edinburgh Science Festival, 
   with projection screen for Actias projection.} 
   \label{fig:Edinburgh}
\end{figure}

The full setup-for a live performance looks as in Figure \ref{fig:Edinburgh}, with a projection of Actias in the background.  \bM One clearly sees the use of two audio-interfaces and two laptops.  We did experiment with a single laptop (and one audio-interface) that both has Actias and Logic Pro installed, but that setup did not work adequately.\footnote{We investigated this together with Moth's technical staff.}  We elaborate further on this issue in the following section.\e

\bM
\section{Restrictions, troubleshooting and solutions}\label{sec:trobleshooting} 

One restriction is that not many guitars with tremolo bar allow for the Fishman pickup.  Besides the  (now discontinued) Fender EOB shown here, another favourite for us are variants of the Ibanez Prestige AZ, where as already mentioned above, instead of the sustainer pick-up we use a Mercury X reverb.  You can see it being used  in video (\ref{eq:merton}), while the Fender EOB setup is being used in video \ref{eq:blackpub}).
Of course, one may not want to use a tremolo bar, and then most guitars allow for the Fishman pickup.  We do feel that the tremolo bar interacts interestingly with the quantum synth, as it enables to produce continuous frequencies more easily than bending strings, that nicely go hand-in-hand with the continuous nature of the quantum synth. The Fishman pickup more than adequately picks up these continuous changes.

Above we already mentioned the need for two laptops and two audio-interfaces for adequate and stable performance.  Of course, this is not needed if one drops the {\bf Mix} layer in Figure \ref{fig:pipeline}, and sends all audio to e.g.~a mixing desk.  For many uses this is evidently possible.  Ourselves, we use Logic Pro as an essential part of our performance.  In that case, having two separate laptops and hence two separate screens is in fact a major bonus anyway.    
\e

Some other technical problems that we encountered with Actias are: 
\bit 
\item Actias sometimes loses its preset, although that never seems to happen when in use.  It always happens  when putting the MIDI-pickup off and then on.  The solution is simple:  don't do that during a performance.
\item Unplugging and replugging MIDI cables sometimes requires deleting an external MIDI device, and introducing it again.  This seems mostly a problem with an Akai MKII  keyboard that we use for testing. 
\item After unmounting and re-mounting the whole setup, Actias needs to be restarted, as otherwise latencies may occur.
\eit
\bM All together, once one is aware of these issues, as we experienced now during extensive studio use, as well as many live performances, the setup is reliable--see video (\ref{eq:merton}).\e

\section{Suggestions for Actias further development}

Rather than only allowing for Z-measurements, it would also be good to have X-measurements as primitive, without the need for rotation to realise them.  This would substantially increase playability, as well as conceptual underpinning.  Also a Y-measurement could be possible.

\bM 
Actias should not restrict to single qubits, but also allow for entangling multiple qubits.  This would enable several musicians to interact with each other in a quantum manner, with in particular, entanglement-induced collapse.  This involves additional variables in order to control the nature of entanglement.  Such a setup has already been performed by ourselves, but then the entanglement, rather than being part of the software, was agreed between musicians.
\e

\section{Some past and planned performances and releases} 

These include:
\bit
\item The launch of Quantum Guitar was  at the Edinburgh Science Festival in 2025, as a keynote talk dedicated to quantum music.
 \item A first performance of ``Bell" -- which uses a new augmented music notation based on ZX-calculus \cite{coecke2023basic} -- in Vienna on the occasion of world quantum day, as part of the research network QISS, and a second performance of ``Bell" at Berlin Medienhaus of the Universitaet der Kuenste (see Figure \ref{fig:Berlin}).
\item Performance of Rakhat-Bi Abdyssagin's ``Quantum Universe" Symphony for Cathedral Organ, Quantum Guitar and Electronics at Merton College Chapel, University of Oxford--\bM see video (\ref{eq:merton}) for the video recording.  This is the first `Quantum Symphony' ever written and performed.\e
\item VIP/Press release of Quantum Guitar at Wacken Open Air festival, August 2026.  Wacken Open Air is the most significant metal festival in the world.
\eit
Our band Black Tish \cite{bigtakeover,youredm} is now recording an album entirely with Quantum Guitar, which is to be released by the end of 2026.  \bM This will include covers/adaptations of Purcell's ``Funeral Music for Queen Mary", and Bob Dylan's ``Masters of War", as well as original tracks in the space of doom metal and post-punk, in order to demonstrate the wide musical applicability of Quantum Guitar.\e

\section{Tech rider} 

Playing the Quantum Guitar requires the following:
\bit
\item a good quality PA for which two XLRs are provided, with good quality stage monitor;
\item a big screen for which an HDMI output is provided, which can either be a projection or a large monitor/television;
\item an armless semi-high chair with adjustable height, and if available, some minor back-support is welcomed, such as in Pictures \ref{fig:Actias} and \ref{fig:Edinburgh});  
\item some good quality drinks.
\eit

\section{Conclusion and outlook}

Quantum Guitar is an instrument that augments an electrical guitar (including tremolo bar) into a quantum musical instrument, and does so without any sacrifice of its `classical features', both in terms of playability and sound.  In particular, it is playable by a single person because no use of hands is required for the quantum enhancement, thanks to using foot controllers.  

Its innovative features:
\bit
\item retaining full classical playability and sound
\item the quantum-enhancement being hands-free
\eit
can be used for also turning other musical instruments  into quantum musical instruments, for example Quantum Violin, Quantum Wind Instrument, Quantum Percussion.  The specific use of specific pedals, which have proven to be reliable, can also be directly copied.

\section{Conflict of interest statement}

We used software of Moth, of which we are a co-founding shareholder.  It is used by my own band, Black Tish (under QBA), and Moth helped with some concerts (see below).  The open-source software will be redeveloped by QEB and possibly others.  

\section{Acknowledgements}

We thank Moth for help with learning to use Actias, and Harry Kumar and Ferdi Tomassini in particular  for assistance with the use of Actias in performances prior to Quantum Guitar, at Wacken Open Air 2024, Lowlands Festival 2024, and at and TUM's International Quantum Forum at Gasteig HP8, Munich, February 2025.

\bibliographystyle{plain}
\bibliography{mainNOWcopy}

\end{document}